\documentclass[manuscript]{aastex62}

\shorttitle{Helioseismic Waves in X9.3 Solar Flare}
\shortauthors{Sharykin et al.}

\begin{document}

\title{Onset of Photospheric Impacts and Helioseismic Waves in X9.3 Solar Flare of September 6, 2017}

\correspondingauthor{Ivan N Sharykin}
\email{sharykin@njit.edu}

\author{Ivan N. Sharykin}
\affiliation{Center for Computational Heliophysics, New Jersey Institute of Technology, Newark, NJ 07102, USA}
\affil{Department of Physics, New Jersey Institute of Technology, Newark, NJ 07102, USA}
\affil{W.W. Hansen Experimental Physics Lab., Stanford University, Stanford, CA 94305, USA}
\affil{Department of Space Plasma Physics, Space Research Institute of RAS, Moscow, 117997, Russia}

\author{Alexander G. Kosovichev}
\affiliation{Center for Computational Heliophysics, New Jersey Institute of Technology, Newark, NJ 07102, USA}
\affiliation{Department of Physics, New Jersey Institute of Technology, Newark, NJ 07102, USA}



\begin{abstract}
The X9.3 flare of September 6, 2017, was the most powerful flare of Solar Cycle 24. It generated strong white-light emission and multiple helioseismic waves (sunquakes).  By using data from Helioseismic and Magnetic Imager (HMI) onboard the Solar Dynamics Observatory (SDO) as well as hard X-ray data from KONUS instrument onboard WIND spacecraft, and Anti-Coincidence System (ACS) onboard the INTERGRAL space observatory, we investigate spatio-temporal dynamics of photospheric emission sources, identify sources of helioseismic waves and compare the flare photospheric dynamics with the hard X-ray (HXR) temporal profiles. The results show that the photospheric flare impacts started to develop in compact regions in close vicinity of the magnetic polarity inversion line (PIL) in the pre-impulsive phase before detection of the HXR emission. The initial photospheric disturbances were localized in the region of strong horizontal magnetic field of the PIL, and, thus, are likely associated with a compact sheared magnetic structure elongated along the PIL. The acoustic egression power maps revealed two primary sources of generation of sunquakes, which were associated with places of the strongest photospheric impacts in the pre-impulsive phase and the early impulsive phase. This can explain the two types of helioseismic waves observed in this flare. Analysis of the high-cadence HMI filtergrams suggests that the flare energy release developed in the form of sequential involvement of compact low-lying magnetic loops that were sheared along the PIL.

\end{abstract}

\section{Introduction}

Energy release of solar flares can affect all layers of the solar atmosphere. The strongest events are accompanied by continuum emission from the photosphere and generation of helioseismic waves. The latter are also referred as ``sunquakes'', initially predicted by \citet{Wolff1972,Kosovichev1995} and discovered by \citet{Kosovichev1998} using Dopplergrams from Michelson Doppler Imager (MDI) onboard the Solar Orbital Heliospheric Observatory (SOHO). Sunquakes are observed in Dopplergrams as concentric waves spreading out from an initial photospheric disturbances occurred during the impulsive phase of a solar flare. The basic information about sunquakes can be found in the reviews of \cite{Donea2011} and \cite{Kosovichev2014}. Helioseismic events are usually associated with appearance of white light emission sources \citep[see statistical work of][]{Buitrago-Casas2015} located close to sunquake sources as found from the helioseismic holography method \citep[][]{Lindsey1997,Donea1999,Lindsey2000}.

There are several ideas about the physical mechanism of sunquake generation. The most popular scenario for initiation of helioseismic waves is a beam-driven hypothesis assuming that accelerated electrons are the primary sunquake driver and reason for the white light emission. In this scenario, the helioseismic waves are formed due to hydrodynamic impact caused by expansion of the chromospheric plasma heated by nonthermal charged particles accelerated in the corona and injected into the chromosphere \citep[][]{Kosovichev1995}. The numerical hydrodynamic modeling of the beam-driven thick-target theory \citep[][]{Kostiuk1975, Livshits1981, Fisher1985, Kosovichev1986, Mariska1989, RubiodaCosta2014} predicts formation of a chromospheric shock (also called ``chromospheric condensation'') moving from the overheated chromospheric plasma into the cooler and denser photosphere. This leads to compression and heating of the photosphere, and generation of the white-light emission and helioseismic acoustic waves. The wave travel through the convective zone where they are reflected and appear as expanding ripples in the photosphere.

However, the plasma momentum can be transferred by other mechanisms, such as a sharp enhancement of the pressure gradient due to eruption of magnetic flux-rope \citep[e.g.][]{Zharkov2011,Zharkov2013} or by an impulse Lorentz force which can be stimulated by changing magnetic fields in the lower solar atmosphere \citep{Hudson2008,Fisher2012,AlvaradoGomez2012,Burtseva2015,Russell2016}. \cite{Sharykin2015a} and \cite{Sharykin2015b} discussed that rapid dissipation of electric currents in the low atmosphere could also explain sunquake initiation. It is possible that different sunquake events are caused by different mechanisms.

To understand the physics of sunquakes and strong photospheric perturbations one needs to observe the whole flare impulsive phase in details to trace the appearance of photospheric impacts. In particular, it is important to study properties of magnetic field in the areas of initial photospheric brightnenings, and also compare with emission sources seen in other parts of electromagnetic spectrum, in particular, with the hard X-ray emission produced by precipitating high-energy electrons.  Observational data with high temporal and spatial resolution are needed to catch the initial photospheric brightnenigs and trace their development in the impulsive and, even, pre-impulsive phases of solar flares.

Recently, \cite{Sharykin2017} used level-1 data from Helioseismic and Magnetic Imager (HMI) onboard Solar Dynamics Observatory (SDO)  \citep{Scherrer2012}, which represent filtergrams taken with different polarization filters across the Fe~I 6173~\AA~ line the time cadence of $\approx 3.6$ s by each of the two HMI cameras. The high temporal resolution allowed them to make a precise comparison between the hard X-ray emission (HXR) observed by Reuven Ramaty High Energy Spectroscopic Solar Imager (RHESSI), photospheric optical emission and sunquake sources of the X1.8 flare of October 23, 2012. It was reported that the initial photospheric emission sources were located in vicinity of the magnetic field polarity inversion line (PIL), and that the time delay between the HXR and photospheric emission profiles did not exceed 4 seconds. This delay was consistent with predictions of the flare hydrodynamics RADYN models. However, the data indicated that the photospheric impact and helioseismic wave might be caused by the electron energy flux, which is substantially higher than that in the current flare radiative hydrodynamic models.

In this paper, we present analysis of X9.3 GOES class solar flare of September 6, 2017, started approximately at 11:53:00 UT, and show that the initial photospheric impacts occurred in the flare pre-impulsive phase. This cannot be explained the standard thick-target flare model. This flare is so far the strongest event of 24 solar cycle, produced protons and lead to Ground Level Enhancement (GLE~72). It was located in active region NOAA~12673 with heliographic coordinates S09W42. The flare generated strong white light emission and helioseismic waves traveling from a large scale photospheric disturbances well seen in all HMI observables. The helioseismic response of this flare were first detected by \citet{Kosovichev2017} who noted an unusual feature: excitation of several sunquakes, probably, by different mechanisms. This flare was located not far from the disk center, and in this case the sunquake signal on HMI Dopplergrams is not reduced due to projection effects, unlike in the X1.8 flare which was near the solar limb \citep{Sharykin2017}.

The main scope of this work is to perform a detailed study of the photospheric impacts which produced strong sunquakes and white light emission, by using HMI data including the standard level-2 HMI observables and the high-cadence level-1 HMI filtergrams, as well as the X-ray data that were available for this flare from the GOES satellite,  the HXR/gamma-ray spectrometer KONUS \citep[][]{Aptekar1995} onboard WIND spacecraft, and, also, from Anti-Coincidence System (ACS) onboard INTERGRAL (SPI) observatory \citep[][]{Vedrenne2003}.  Our first task is to trace dynamics of photospheric emission sources relative to the magnetic field structure in the flare region, and, thus, to define places of the initial photospheric impacts. The second task is to compare spatial positions of the photospheric impacts seen in different HMI observables with the sunquake sources deduced from the helioseismic holography method. The third task is to compare time profiles of the photospheric emission sources (using HMI filtergrams) with the HXR time profiles in order to test the beam-driven hypothesis of photospheric flare perturbations.

Section~\ref{Sec_PH} describes the relationship between the magnetic field structure of the flare region and impulsive sources of continuum emission and Doppler shift using HMI 45-second level-2 data. The temporal behavior of the sources is compared with the corresponding GOES soft X-ray lightcurves, as well as with the HXR time profiles from KONUS/WIND and ACS/INTEGRAL. Section~\ref{Sec_SQ} is devoted to analysis of helioseismic signals (sunquakes) from the flare region. It presents time-distance analysis of the observed helioseismic waves and reconstruction of sunquake sources using the acoustic holography method. Section~\ref{Sec_lev1} presents analysis of photospheric emissions using the HMI level-1 filtergrams to determine the precise timing of the photospheric impacts relative to the flare HXR signals from KONUS/WIND. The last section summarizes results and formulates conclusions.

\section{Photospheric Impacts and X-ray Emission}\label{Sec_PH}

 \subsection{Polarity Inversion Line and Distribution of Continuum Emission and Doppler-shift Sources}

\begin{figure}
\centering
\includegraphics[width=1.0\linewidth]{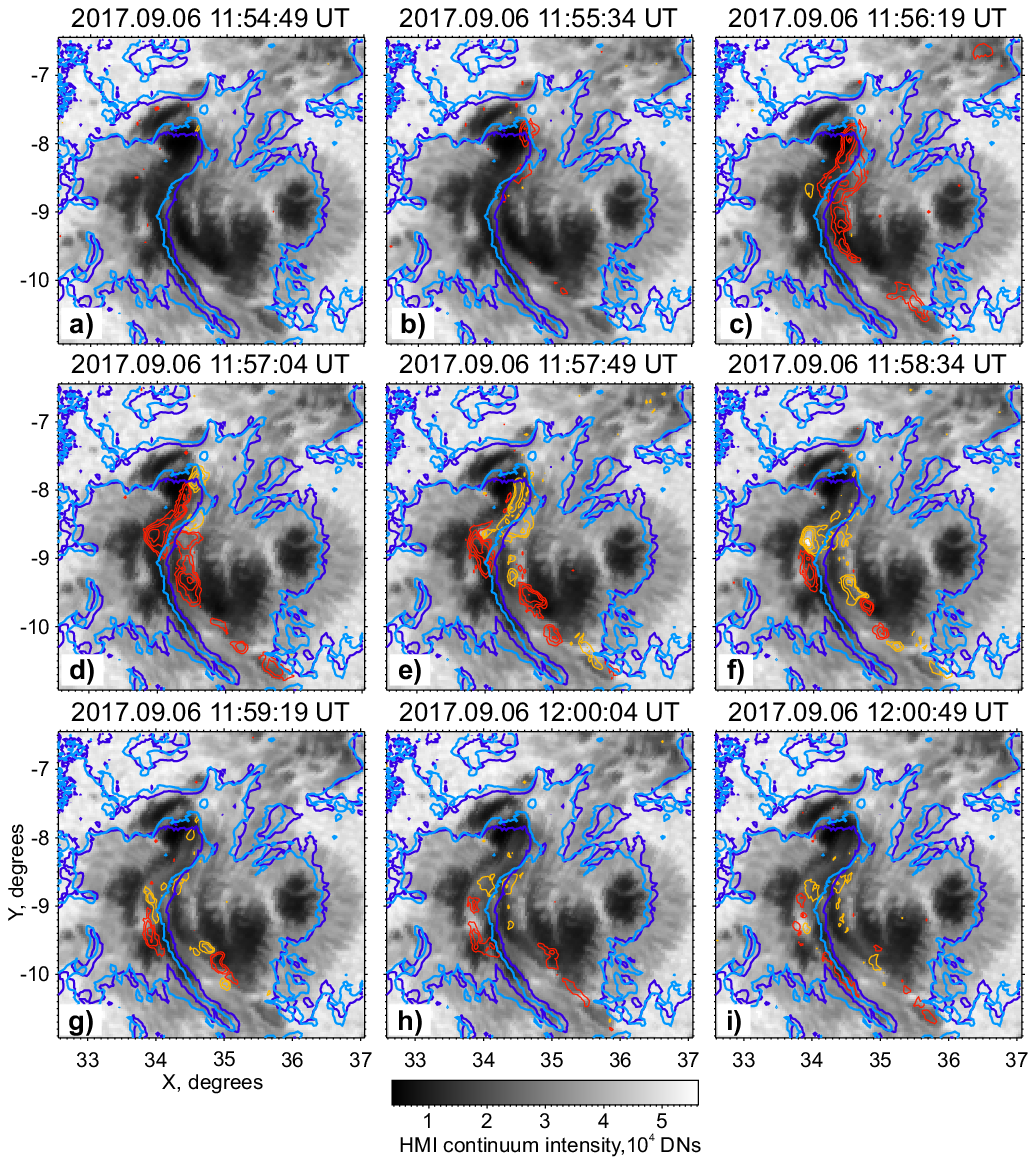}
\caption{A series of HMI continuum intensity maps (black-white background images) projected onto the Heliographic coordinates, showing the flare pre-impulsive and impulsive phases. Red and orange contours correspond to positive and negative changes of the running time differences of the HMI intensity maps with levels of 2.5, 5, 10, and 20 kDNs. Blue and cyan lines show the PIL from the HMI LOS magnetograms for two time moments: before (11:30:04~UT) and after (12:59:19~UT) flare.}
\label{HMI_Ic}
\end{figure}

The X9.3 flare of September 6, 2017, was characterized by very strong photospheric impacts seen in all HMI observables. In this section we analyze the line-of-sight (LOS) magnetograms, Dopplergrams,  and continuum intensity maps taken by the HMI instrument with cadence of 45 seconds and spatial resolution of 1$^{\prime\prime}$.  Figure\,\ref{HMI_Ic} presents a sequence of the continuum intensity maps covering the pre-impulsive and impulsive phase. These maps are reprojected onto the heliographic grid to remove the projection effect and demonstrate the true length scale of the flare impacts and their position relative to the magnetic field polarity inversion line (PIL) from the LOS Magnetograms. The flare produced very strong perturbations that distorted the magnetograms and the PIL's shape. So, the two PILs for preflare (blue curves) and postflare (cyan) times are presented in Fig.\,\ref{HMI_Ic}.

One can see that the flare region is very complex. Several sunspots are located close to each other forming a $\delta$-type configuration with a S-shaped PIL. The photospheric emission sources determined from the running difference of the continuum intensity maps are marked by red and orange contours. Red and orange colors correspond to positive and negative frame-to-frame changes, respectively. Images in Fig.\,\ref{HMI_Ic}a-c correspond to the pre-impulsive phase (before the start of HXR emission). One can notice in panel $b$ that the initial perturbations develop from two brightenings located in the PIL. In the next 45~s (panel $c$), the perturbations expanded along the PIL in the form of two sheared flare ribbons on both sides of the PIL. Additional brightnings appeared in the southern part of the large sunspots. The largest spatial scale (distance between the southern and northern remote sources) is about 28~Mm. The distance between the initial brightnengs along the PIL is about 10~Mm, and the distance between the photospheric ribbons across the PIL is 3~Mm. These observations suggest small-scale sheared magnetic loops in the PIL region were activated during the initial pre-impulsive energy release with subsequent involvement of a large-scale magnetic structure. It is also worth noting that the flare ribbons shown in panel $c$ are very structured with many emission cores. The 45-sec HMI data allow us to separate the photospheric flare sources in time and space. However, analysis of high-cadence HMI filtergrams in the Section~\ref{Sec_lev1} will show more details.

During the impulsive phase (Fig.\,\ref{HMI_Ic}d-i), the emission sources moved along the PIL in the southern direction, probably, reflecting involvement of new magnetic loops into the flare energy release process. At the same time, continuum emission of the initial brightenings started to decrease (marked by orange contours).

\begin{figure}
\centering
\includegraphics[width=1.0\linewidth]{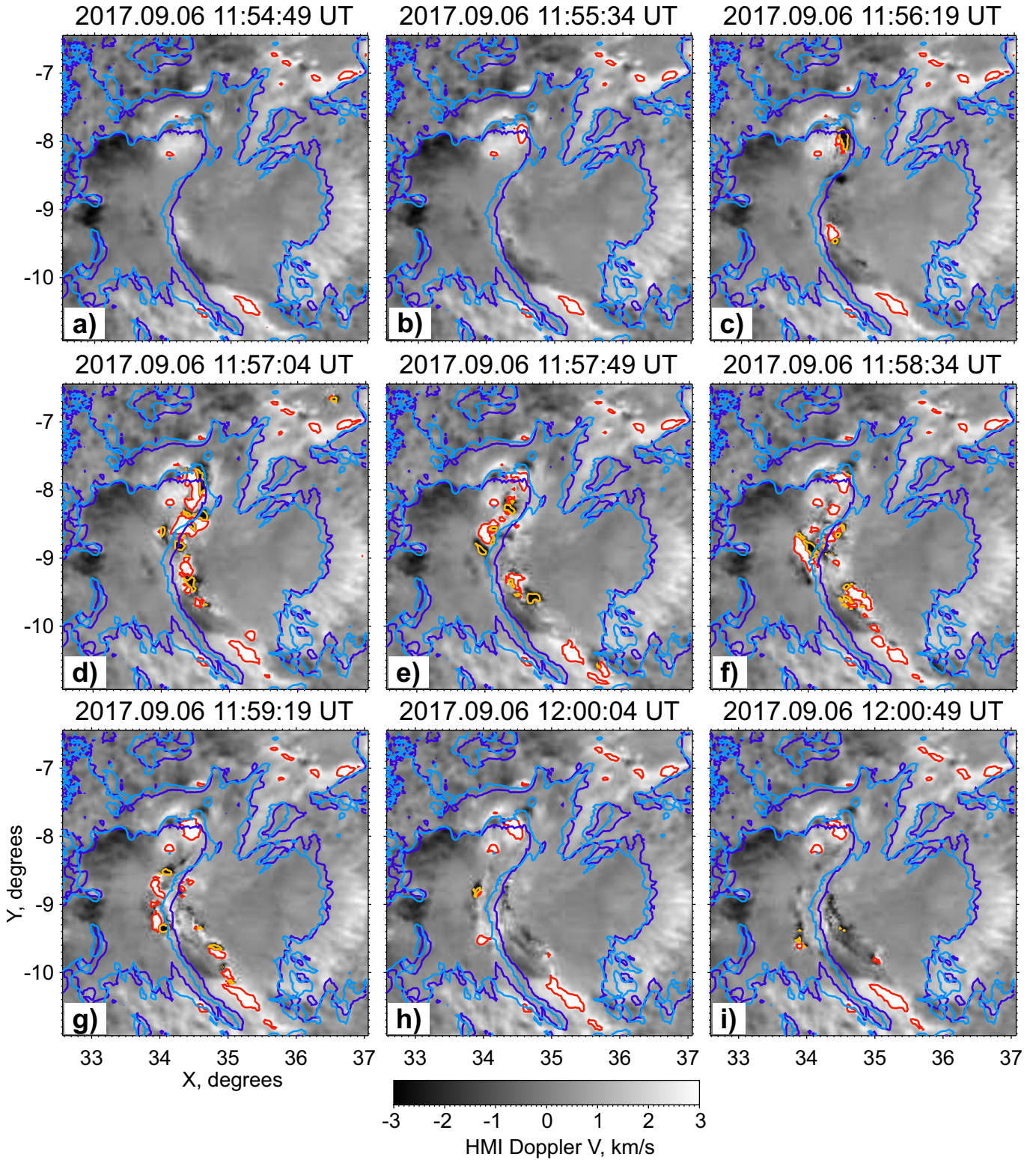}
\caption{A series of HMI Dopplergrams (black-white background images) projected onto the Heliographic coordinates for the same moments as in Fig.~\ref{HMI_Ic}. Red and orange contours correspond to positive and negative Doppler velocities with magnitude of 3~km/s. Blue and cyan lines show the PIL from HMI LOS magnetograms for two time moments: before (11:30:04~UT) and after (12:59:19~UT) flare.}
\label{HMI_V}
\end{figure}

Figure\,\ref{HMI_V} demonstrates the HMI Dopplergrams for the same time moments as in Fig.\,\ref{HMI_Ic}. The strong photospheric impacts are revealed as compact white and black patches.  They are also highlighted by red and blue contours corresponding to positive (downward) and negative (upward) Doppler velocities with the magnitude of 3~km/s. The highest velocity magnitudes were up to 14 km/s and 8 km/s for downward and upward velocities, respectively. However, these values may not accurately characterize the real plasma velocities. Because of strong distortion of the Fe~I line profile (observed by HMI)  the Doppler shift measurements in such places may be incorrect. Nevertheless, the high Doppler-shift values indicate places of strong photospheric impacts, and help  to detect sources of helioseismic waves \citep[e.g.][]{Kosovichev2014}. The general Doppler velocity response was mostly downward that is in accordance with the idea of a downward moving shock producing the photospheric impact and helioseismic waves.

In the pre-impulsive phase (Fig.\,\ref{HMI_Ic}a-c), the initial velocity perturbations were located very close to the PIL. Comparing Fig.\,\ref{HMI_V} and Fig.\,\ref{HMI_Ic} frame by frame one can find that the continuum intensity changes correspond to the Doppler velocity impacts. However it should be kept in mind that we compare time differences (for continuum intensity) with the HMI Dopplergrams. However, there are some differences between the Dopplergrams and intensity difference maps (which partly may be due the 22.5~sec time difference). For example, in the pre-impulsive phase (panels $c$) the Doppler perturbations were located at the edges of the continuum intensity ribbons. Panels $d$ (beginning of the impulsive phase) revealed that the photospheric impacts deduced from the Dopplergrams were located closer to the PIL than the intensity perturbations. Moreover the strongest impact was located directly in the PIL according to the Dopplergram. The subsequent frames $e-i$ from both data sets show similar strongest photospheric impacts.

    \subsection{Comparison of the Photospheric Signals and HXR Time Profiles}

\begin{figure}
\centering
\includegraphics[width=0.6\linewidth]{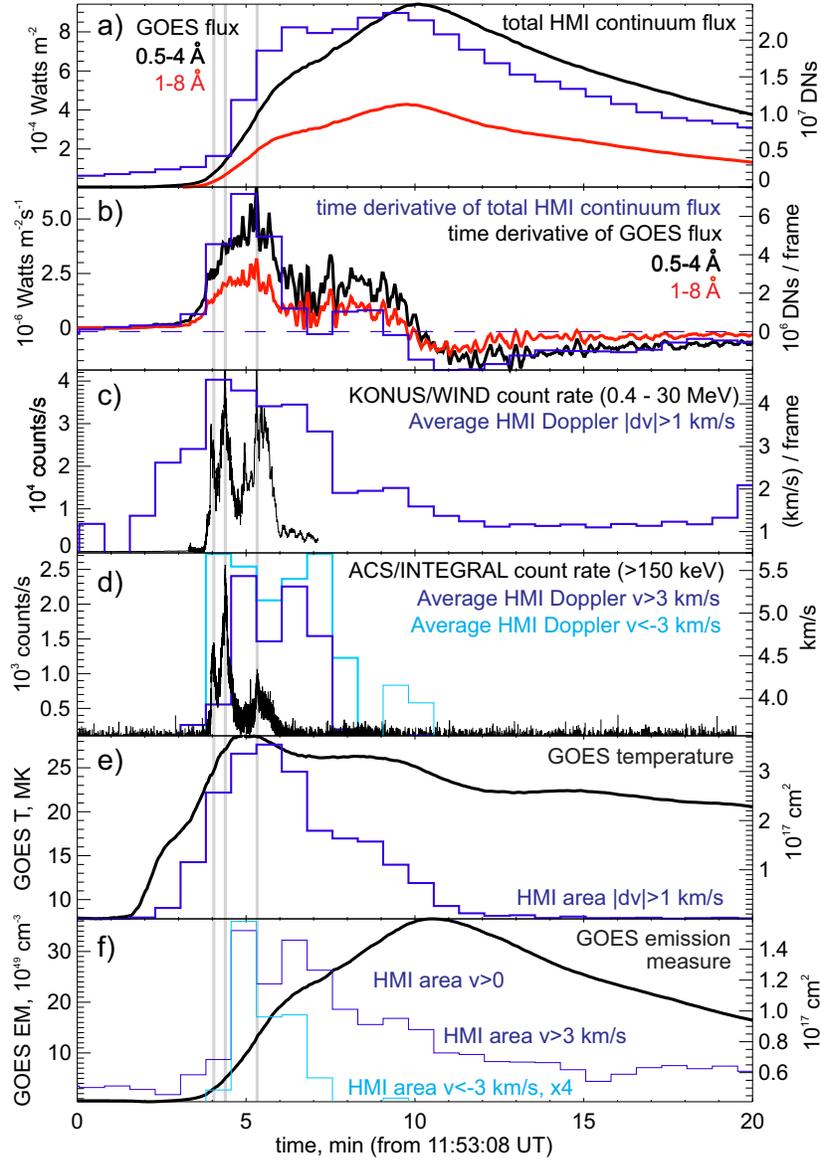}
\caption{a) The total HMI continuum flux (blue) from the flare region (integrated over FOV of images in Fig.~\ref{HMI_Ic}) as a function of time and the soft X-ray lightcurves in two GOES channels of 0.5-4 (red) and 1-8 \AA{} (black); b) Time derivatives of the total HMI continuum flux and GOES lightcurves; c) Time profiles of KONUS/WIND count rate in the energy range of 0.4-30 MeV (black) and the mean HMI Doppler velocity calculated for regions with $|dv|$$>1$~km/s (blue); d) The ACS/INTEGRAL HXR data (above 150~keV) and the mean HMI Doppler velocity (with amplitudes higher than 3 km/s). Blue and cyan colors correspond to positive and negative Doppler velocities, respectively; e) Temporal profiles of the total area of regions with Doppler velocity variations higher than 1~km/s (blue) and the flare temperature calculated from the GOES data (black); f) Temporal profile of the total area of regions with the Doppler velocity amplitude higher than 3~km/s (blue and cyan lines) and the flare emission measure calculated from the GOES data.}
\label{TPs}
\end{figure}

In this subsection we compare of HMI flare signals averaged through field-of-view (FOV) in Figures~\ref{HMI_Ic} and~\ref{HMI_V} with the soft X-ray (SXR) and the hard X-ray (HXR) data. Figure\,\ref{TPs} shows comparison between the HMI observables (shown as step-wise functions)  and the SXR and the HXR time profiles.

It follows from Fig.\,\ref{TPs}a that the total HMI continuum intensity flux varies in accordance with the lightcurves of the GOES channels 0.5-4 and 1-8~\AA{}. The time derivatives (Fig.\,\ref{TPs}b) also fit each other. The long duration of  the photospheric continuum and SXR emissions indicates that they are of a thermal origin. However, the SXR emission comes from the hot coronal plasma while the HMI continuum emission can come only from the relatively low-temperature photospheric plasma. The temperature and emission measure of the SXR emitting plasma (calculated in the single-temperature approach) are plotted in panels $e$ and $f$, respectively. The highest temperature of 29~MK was during the impulsive phase when the peak emission measure was about $3.6\times 10^{50}$~cm$^{-3}$.

The impulsive phase of this flare was not observed by Reuven Ramaty High-Energy Spectroscpic Solar Imager (RHESSI) and FERMI spacecrafts. Therefore, we used the HXR data from the HXR/gamma-ray spectrometer KONUS \citep[][]{Aptekar1995} onboard WIND spacecraft, and, also, from Anti-Coincidence System (ACS) onboard the INTERGRAL (SPI) space observatory \citep[][]{Vedrenne2003}.

The KONUS/WIND is a experiment devoted to study gamma-ray bursts and solar flares. It consists of two NaI(Tl) detectors observing correspondingly the opposite celestial hemispheres and sensitive to all incoming HXR and gamma-ray emissions. The instrument operates near Lagrange point L1, so it does not suffer from ``nights'', and has a very stable background. The instrument works in two modes: waiting and triggered mode. In the first mode, the count rate light curves are available in three wide energy channels G1 (18-70 keV), G2 (70-300 keV), G3 (300-1160 keV) with time cadence of 2.944 s. Switching to the triggered mode occurs at a statistically significant excess of ~9 sigma above background in the G2 energy channel. In the triggered mode the count rates are measured in the same three channels with a varying time resolution from 2 to 256 ms and with the total duration of 250 s. The KONUS/WIND data have been used in application to solar flares  \citep[e.g.][]{Fleishman2016,Lysenko2018}. The HXR count rate from KONUS/WIND in G2 channel for our flare is plotted in Fig.\,\ref{TPs}c.

The ACS instrument is a set of BGO crystals viewed by photo-multipliers. Due to the large effective area of 0.3 m$^2$ the ACS is a very sensitive instrument to all HXR/gamma-ray emissions in the energy range $\gtrsim$100 keV coming from all directions. The time resolution of the ACS is 50~ms. The SPI satellite has a very eccentric orbit outside the radiation belts. Thus, the radiation background is also quite stable on flare time scales. The ACS is not a solar-dedicated instrument, but its data have been used for solar flare studies \citep[e.g.][]{Struminsky2010, Zimovets2012}. The HXR count rate from ACS is shown in Fig.\,\ref{TPs}d.

The HXR data from both instruments are compared with the integral Doppler velocity responses in Fig.\,\ref{TPs}c-d. The HXR time profiles are characterized by three major impulses the approximate duration of which was 30-50 seconds. An addition, the time profiles reveal fine temporal structure of this impulses. Total duration of the HXR phase was about 150~seconds. Comparing the peak magnitudes in the different energy ranges we conclude that the first two peaks were harder in terms of the X-ray emission spectrum than the third one which corresponds to the peak of time derivatives of the SXR lightcurves from GOES.

In this respect it is worth noting that the SXR and HXR time profiles do not follow the \citet{Neupert1968} law. This may be related to the fact that we consider the high energy data, above 70~keV. For lower energies there might be better correspondence.

The piece-wise plot in Fig.\,\ref{TPs}c demonstrates temporal evolution of the averaged Doppler velocity changes (using the running time differences of Dopplergrams) for the region where the absolute value of velocity variations were higher than 1~km/s. One can see that the Doppler velocity perturbations were already pronounced (up to 3 km/s) in the pre-impulsive phase during the first three HMI frames before the first HXR pulse. Then, the highest Doppler velocity was achieved during the first two HXR pulses. However, the largest area (Fig.\,\ref{TPs}e) of regions with the Doppler velocity changes higher than 1~km/s was achieved at the time of the softer third HXR pulse.

The average Doppler velocity (Fig.\,\ref{TPs}d) was calculated by averaging over the regions where the absolute values of velocity were higher than 3~km/s. Both, the downward and upward velocities were most intensive also around the first two HXR peaks. The same temporal dynamics was observed for the Doppler perturbation area (Fig.\,\ref{TPs}e). However, strong Dopplergram perturbations were also observed after the third HXR peak.

From Fig.\,\ref{TPs}d we also found that the plasma temperature deduced from the GOES data had a peak around the HXR peaks, corresponding to the time derivatives shown in panel Fig.\,\ref{TPs}b. Thus, the coronal plasma heating was simultaneous with the precipitation of nonthermal electrons into the dense solar atmosphere and the rise of the photospheric continuum emission.


\section{Helioseismic Flare Signals}
\label{Sec_SQ}

To analyze helioseismic waves and their sources we used time series of running differences of HMI Dopplergrams remapped onto the heliographic  coordinates and tracked with solar rotation. 
To isolate the wave signal from convective noise we applied a Gaussian frequency filter with a central frequency of 6 mHz and width of 2 mHz to each pixel of the Dopplergram differences.

\begin{figure}
\centering
\includegraphics[width=1.0\linewidth]{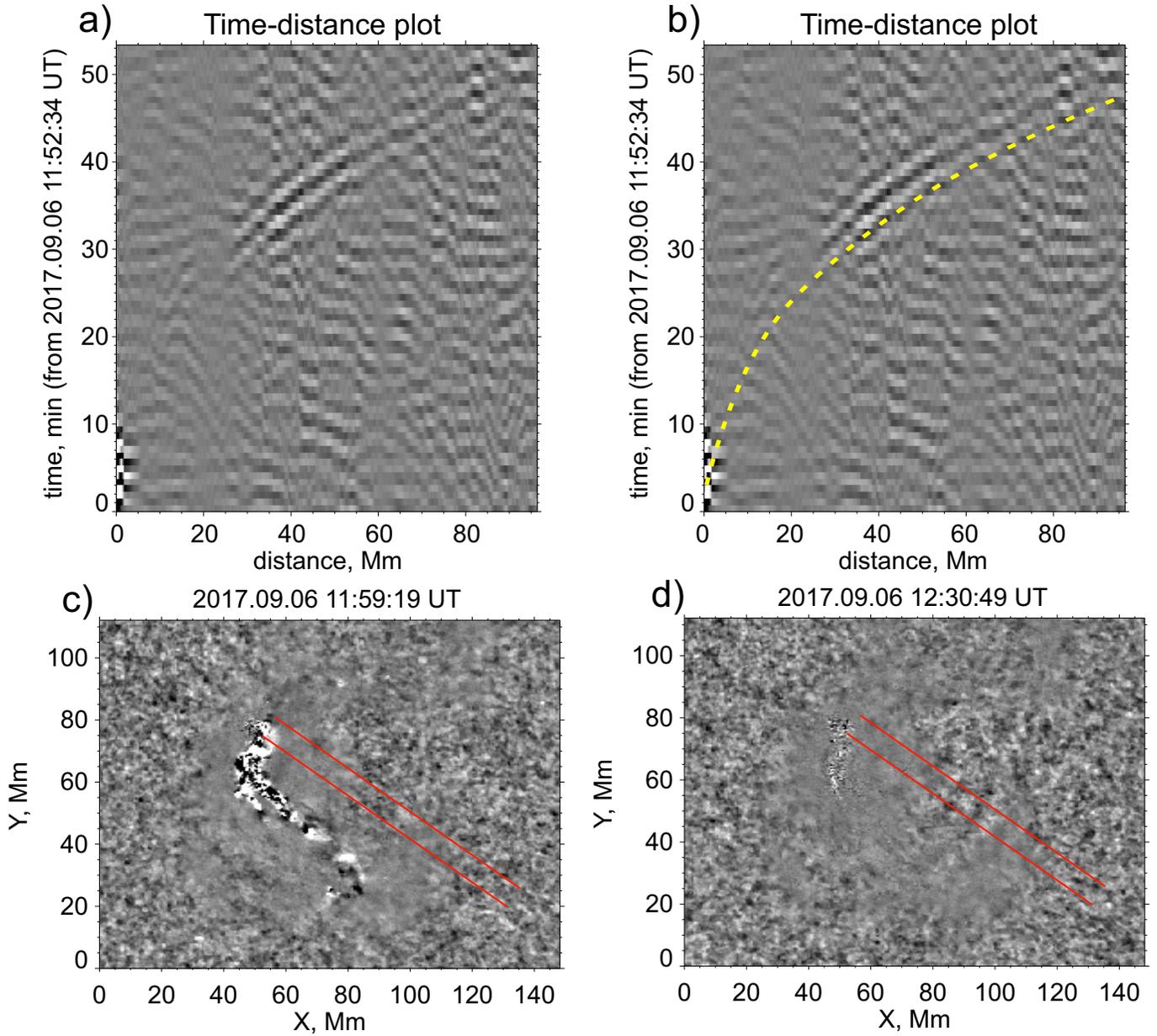}
\caption{a-b) The sunquake time-distance diagram calculated along the red lines plotted in panel~c and d. c-d) the time differences of Dopplergrams projected onto the Heliographic coordinates and filtered with a Gaussian frequency filter centered around 6 mHz for two moments of time showing the photospheric impact (c), and the helioseismic wave front (d).}
\label{SQ_TD}
\end{figure}

The sunquake event was observed as waves of circular shape spreading out from the region of the flare Dopplergram disturbances described in the previous section. Example of flare disturbance and the sunquake wave is shown in Fig.\,\ref{SQ_TD}c,d. The time-distance (TD) diagram for this wave is shown in Fig.\,\ref{SQ_TD}a. Two red lines show the orientation of the image slice, along which we calculated the diagram by averaging across this slit. The bands were  oriented perpendicular to the wave front, and  were twenty pixels wide to increase the signal-to-noise ratio. The theoretical time-distance relation calculated in the ray approximation for a standard solar interior model of \cite{Christensen-Dalsgaard1993} is marked by dashed curve in the TD~diagram in Fig.\,\ref{SQ_TD}b. The position of the wave ripples in the TD diagram matches the theoretical model. Thus, the observed wave was generated in the source corresponding to the Dopplergram disturbance.

\begin{figure}
\centering
\includegraphics[width=0.82\linewidth]{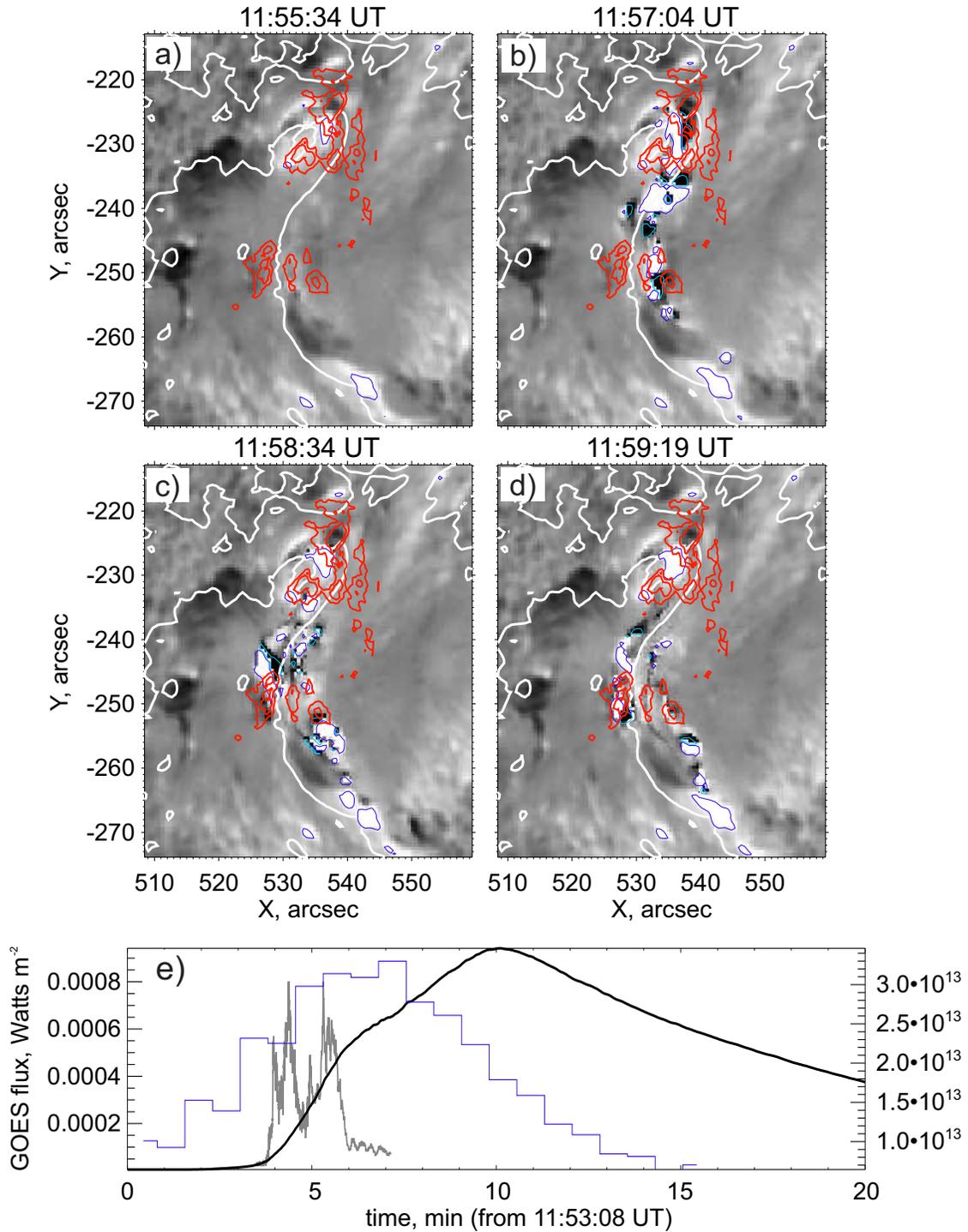}
\caption{a-d) Comparison of the total acoustic egression power map (red contours with 30 and 50\% levels) with the flare Dopplergrams (black-white background images) for different time moments. Blue and cyan contours correspond to positive and negative Doppler velocities with levels of 3~km/s. White line marks the PIL deduced from the HMI vector magnetogram for time moment 12:12:04~UT; e) Time profiles of the total egression power (blue), the GOES lightcurve in band of 1-8~\AA{} (black) and the KONUS/WIND count rate in the energy range of 0.4-30~MeV (gray).}
\label{Egress}
\end{figure}

To reconstruct the two-dimensional structure of the seismic source we used the helioseismic holography method \citep{Lindsey1997,Donea1999,Lindsey2000}. This approach uses a theoretical Green function of helioseismic waves to calculate the egression acoustic power corresponding to the Doopler velocity perturbations. The egression acoustic power map made in the frequency range of 5-7 mHz is shown in Fig.\,\ref{Egress}a-d by red contours. We calculated this map by summing the egression acoustic power snapshots within the time interval found from the uncertainty principle $\Delta t\sim 1/\Delta \nu\approx 500$ seconds, where $\Delta \nu=$2 mHz. This time interval corresponds to the appearance of the strong Doppler velocity perturbations. The egression power map is compared with four different Dopplergrams: two for the pre-impulsive phase (Fig.\,\ref{Egress}a and b)  and two for the impulsive phase (Fig.\,\ref{Egress}c and d). The PIL plotted in these panels was determined for the vertical magnetic field component using the HMI vector magnetogram reprojected onto the heliographic grid.

The egression power map shows a complex distribution of helioseismic sources which were located in the close vicinity to the PIL. There are two regions of generation of the helioseismic waves. Several sources are located in the same place as the initial perturbations observed during the pre-impulsive phase and the first HXR peak (Fig.\,\ref{TPs}). Possibly, the helioseismic waves were generated in the late pre-impulsive phase at the start of the first HXR pulse. However, we cannot confirm this from the acoustic egression map because it is averaged over the whole the impulsive phase. The southern helioseismic sources were located in the place of the photospheric perturbations observed around the second and third HXR peak. Thus, one can conclude that helioseismic waves were generated during the whole impulsive phase and, probably, even in the pre-impulsive phase. Perhaps, the superposition of the helioseismic waves excited during the last two HXR peaks can explain the unusually long wavelength of the south-ward helioseismic wave \citep{Kosovichev2017}.

The temporal profile (Fig.\,\ref{Egress}e) of the egression acoustic power reveals a maximum two minute later than the third HXR peak. The power started to increase in the pre-impulsive phase. However, we can conclude that the most efficient generation of helioseismic waves was definitely during the precipitation of nonthermal electrons into the lower layers of solar atmosphere.

\section{Precise Timing of Photospheric Impacts from HMI Level-1 Data}
\label{Sec_lev1}

The HMI produces data by scanning the magneto-sensitive Fe~I line (6173~\AA) at six wavelength positions across the line profile \citep{Couvidat2016}. There are two cameras producing a series of filtergrams with the pixel size of 0.5$^{\prime\prime}$ in linear polarization (Camera 1) and in right and left circular polarizations (Camera 2). The filtergrams from both cameras (level-1 data) are used to reconstruct the full Stokes profiles. To calculate the line-of-sight magnetograms, Dopplergrams and continuum intensity (level-2 data), only Camera 2 is used. We use the original filtergram data in the disk coordinates from both cameras to achieve a high temporal resolution in order to investigate dynamics of the photospheric flare emission sources. The time cadence of filtergrams from each cameras is 3.6 seconds. Previously, HMI filtergrams were applied to study a limb flare in the works of \citet{SaintHilaire2014} and \citet{MartinezOliveros2014}. \citet{Sharykin2017} presented analysis of the X-class flare of October 23, 2012, which produced strong sunquakes. A detailed description of how to use the filtergrams for analysis of photospheric emission sources with high temporal resolution can be found in their paper. Here we present only a brief description of the filtergram analysis technique.

\begin{figure}
\centering
\includegraphics[width=1.0\linewidth]{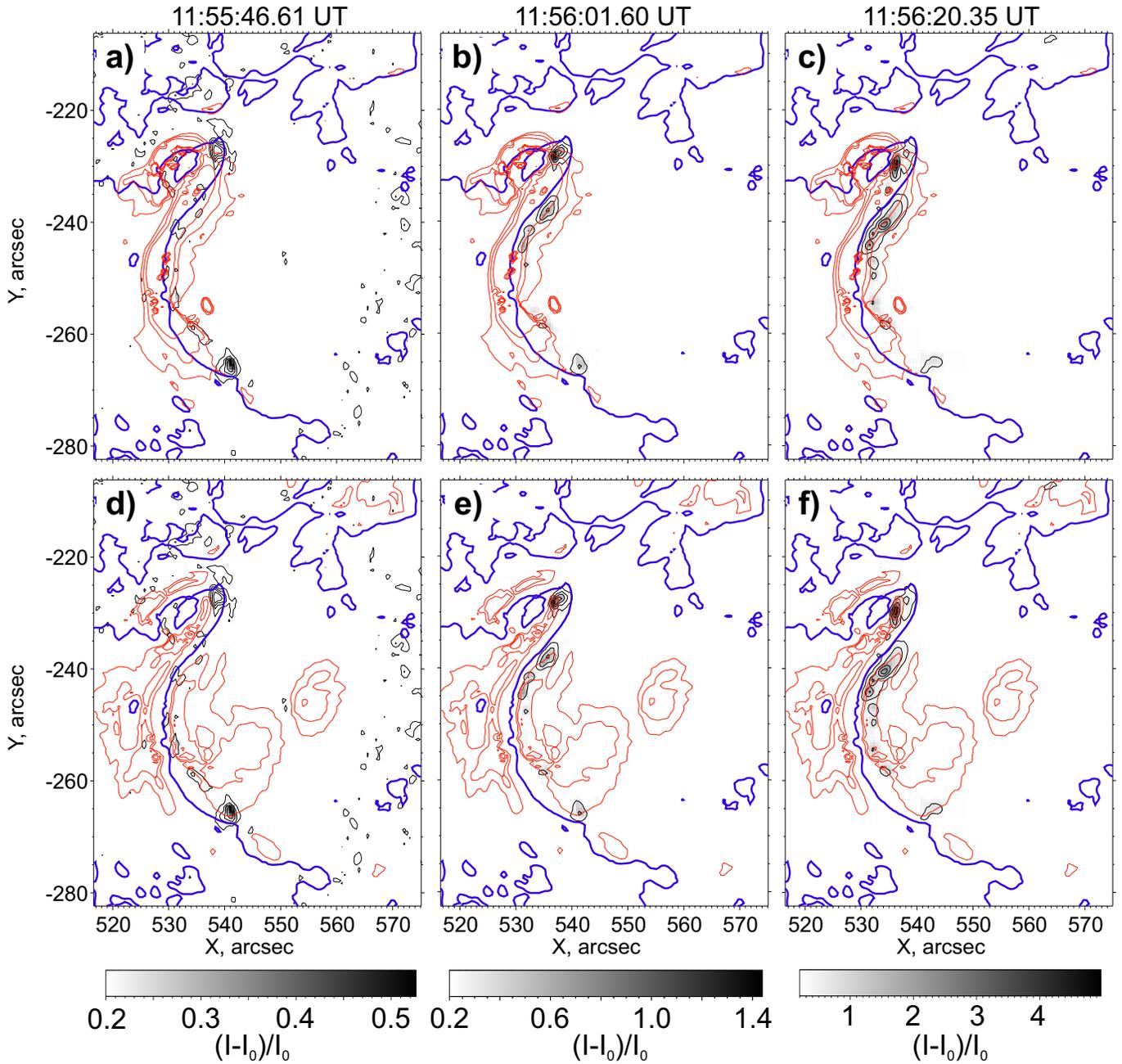}
\caption{Photospheric emission maps (black-white background images) in the pre-impulsive phase, determined from the HMI level~1 data for three moments of time are compared with the horizontal (red contours in panels a-c) and vertical (red contours in panels d-f) magnetic fields. Contour levels correspond to 1.5, 2, and 2.5~kG. The PIL is marked by blue color.}
\label{lev1_preimp}
\end{figure}

\begin{figure}
\centering
\includegraphics[width=1.0\linewidth]{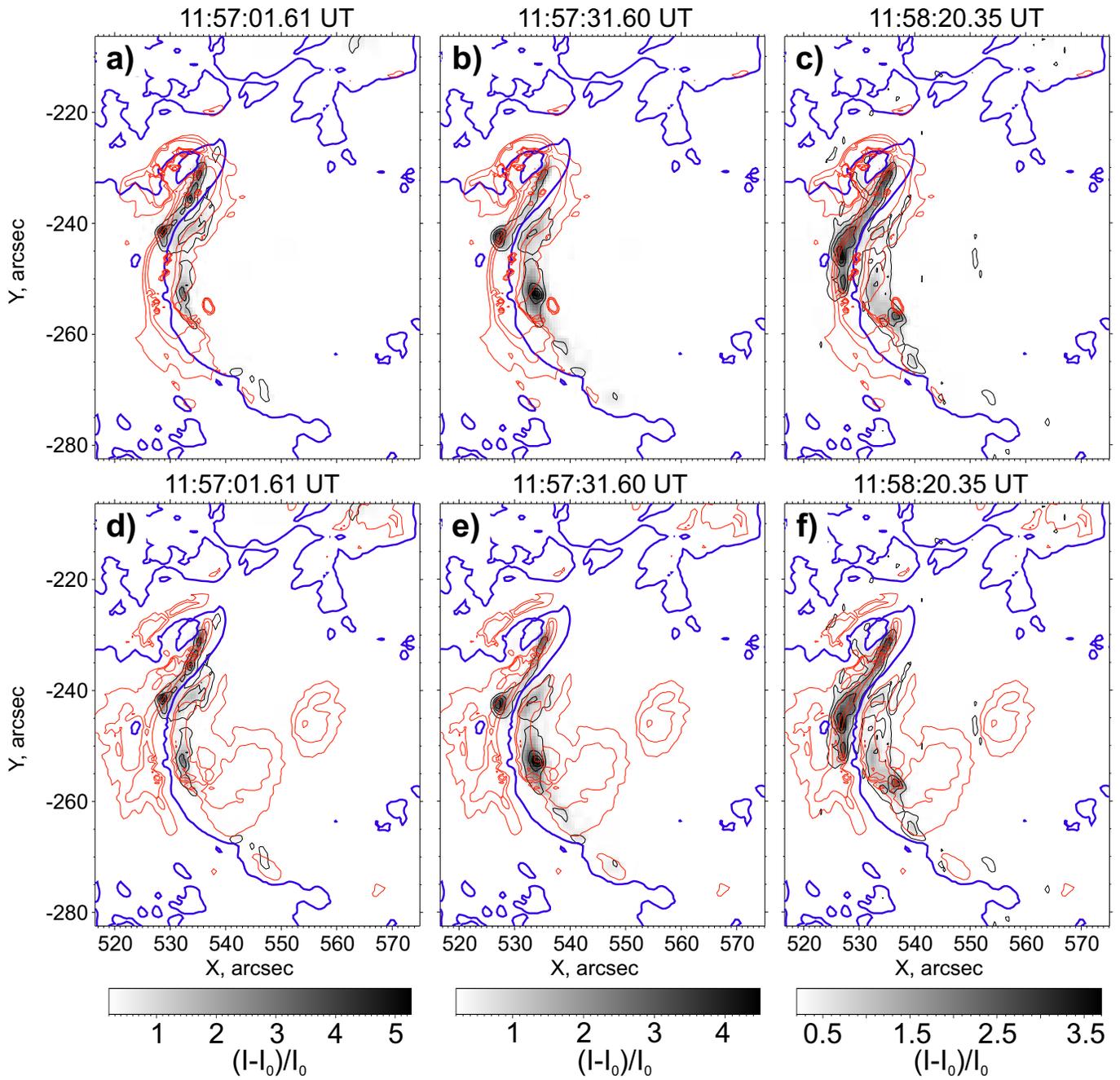}
\caption{The same as in Fig.~\ref{lev1_preimp} for three moments during the flare impulsive phase.}
\label{lev1_imp}
\end{figure}

To identify the flare signals using the HMI level-1 data we subtract a preflare filtergram from the filtergrams taken during the impulsive phase for the same wavelength and polarization. This allows us to detect changes in the flare region with high temporal resolution. The value in each pixel is calculated as $(I-I_0)/I_0$, where $I$ is the pixel value for a flare filtergram, and $I_0$ is the corresponding preflare value. We decided not to perform frequency filtering of the filtergram lightcurves to remove the variations caused by the line scanning  as it was done in the paper of \citet{Sharykin2017} because  the photospheric signal was sufficiently strong in the flare emission sources. To demonstrate development of the photospheric emission we plot three filtergrams for 35~sec of the pre-impulsive phase (Fig.\,\ref{lev1_preimp}) and three filtergrams for 80~sec of the impulsive phase (Fig.\,\ref{lev1_imp}), and compare these with the maps of the horizontal and vertical magnetic field strength. To highlight the emission sources we also plotted black contours with 20, 40, 60, and 80\% levels. Red contours in the top panels (a-c) of both figures mark the horizontal magnetic field strength of 1.5, 2, and 2.5~kG. The same levels of the vertical magnetic field are shown by red contours in the bottom (d-f) panels. The PIL is shown by blue lines. The magnetic field maps are derived from the HMI vector magnetograms with the time cadence of 720 seconds and, thus, the red and blue contours are the same all panels. Only the black contours showing the continuum emission sources change during the selected time intervals.


The first photospheric emission in the pre-impulsive phase was generated from two compact distant sources (Fig.\,\ref{lev1_preimp}a) in the PIL. It seems that these sources are associated with a magnetic structure elongated along the PIL. The subsequent filtergrams (Fig.\,\ref{lev1_preimp}b,c) reveal emission in the PIL from a region of a very strong magnetic field (up to 3~kG) which is mostly horizontal. Positions of the strongest photospheric disturbances in the pre-impulsive phase correspond to the northern region of sunquake generation.
It is worth noting that the observed photospheric emission comes from numerous brightenings in the PIL, probably, due to fragmented flare energy release.

Transition from the pre-impulsive phase to the impulsive phase is clearly seen from the HMI filtergrams (Fig.\,\ref{lev1_imp}a) as appearance of new brightenings along the PIL, south relative to the initial impacts. The strongest emission sources in Fig.\,\ref{lev1_imp}b,c are located about five arcsec from the PIL where the magnetic field is predominantly vertical. The location of the strongest brightenings is close to the southern complex of acoustic egression sources.

\begin{figure}
\centering
\includegraphics[width=1.0\linewidth]{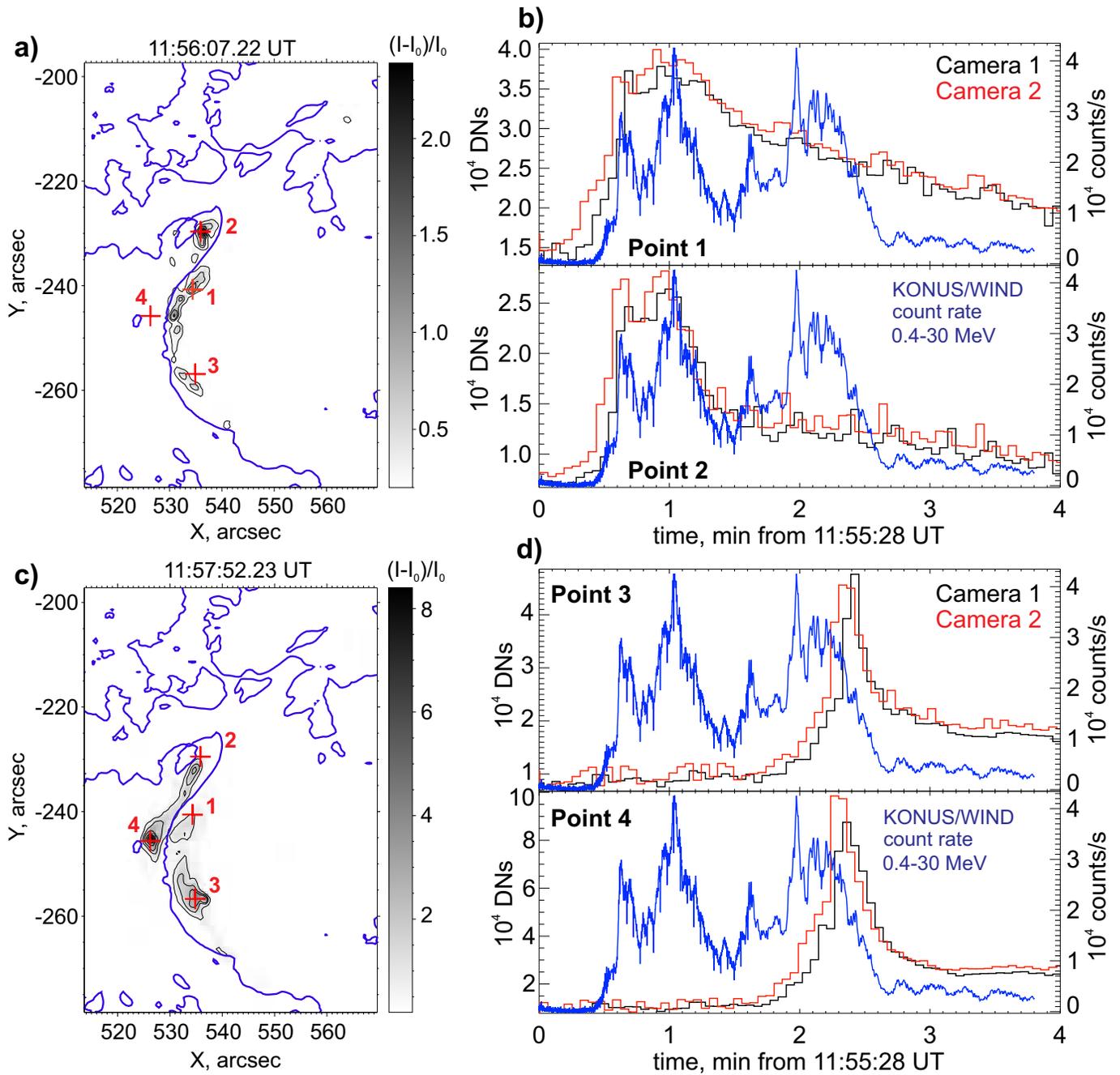}
\caption{a and c) The flare emission maps from the HMI filtergrams for two moments of time during the impulsive phase. The PIL is marked by blue color; b and d) The photospheric emission lightcurves (black for HMI Camera 1 and red for Camera 2) at four different points marked by red crosses in panels a and c, and the KONUS/WIND count rate (blue) in the energy range of 0.4-30~MeV measured from the whole Sun. }
\label{TPs_lev1}
\end{figure}

In Figure \ref{TPs_lev1} we compare the filtergram lightcurves from four characteristic flare points (shown by red crosses in panels a and c) with the HXR time profiles. These points were selected to characterize the dynamics of continuum emission in the vicinity of the photospheric impacts observed during the pre-impulsive (points 1 and 2, panel b) and impulsive (points 3 and 4, panel d) phases. The filtergrams corresponding to the pre-impulsive and the impulsive phase are shown in panels a and c, respectively.  The resulted lightcurves are compared with the KONUS/WIND HXR time profile (blue line) in the energy range of 70-300~keV. This comparison reveals that the photospheric emission during the first and second HXR peaks were associated with points 1 and 2, appeared during the pre-impulsive phase. Points 3 and 4 generated emission during the last HXR pulse. Thus, the flare continuum emission was sequentially generated along the PIL starting from the places of strong horizontal magnetic field in the pre-impulsive phase and finishing in sunspot areas of predominantly vertical magnetic field.

From the analysis of the HMI filtergrams one can conclude that activity in the pre-impulsive phase and initial impulsive phase was associated with energy release in a compact sheared magnetic structure elongated along the PIL with strong horizontal magnetic field. Subsequently, the energy release occurred in a larger magnetic structure sheared along the PIL and the footpoints located in areas of predominantly vertical magnetic field. The observed emission sources are very dynamic, compact and fragmented. Two main flare regions producing sunquakes were associated with the photospheric emission sources seen in the pre-impulsive and impulsive phase, respectively. This can explain the two types of helioseismic waves detected by \citet{Kosovichev2017} in this flare, but the excitation mechanism is still unknown. It requires high-resolution spectro-polarimetric data with cadence higher than it is currently available from HMI to determine the physical conditions in photospheric emission sources.

\section{Discussion and Conclusions}

We presented a study of the X9.3 solar flare (SOL2017-09-06), which revealed strong white light emission and generation of helioseismic waves (sunquakes). This study focused on three tasks: 1) investigate spatio-temporal dynamics of photospheric emission sources; 2) identify sources of helioseismic waves and compare them with the photospheric impacts; and 3) compare dynamics of photospheric emission with hard X-ray temporal profiles. To perform these tasks, in addition to the standard (level-2) HMI observables obtained with 45 sec and 135 sec cadence we used high-cadence (3.6 s) HMI filtergrams (level-1 data). This allowed us to localize initial photospheric brightenings, determine their relationship to the HXR pulses observed with high temporal resolution by the KONUS/WIND and ACS/INTEGRAL instruments, and relate the energy release events to sources of the helioseismic waves.

The main results can be summarized as following:

\begin{enumerate}

\item The photospheric flare impacts started to develop in compact regions in close vicinity of the magnetic polarity inversion line (PIL) in the pre-impulsive phase before detection of the HXR emission. The initial photospheric disturbances were localized in the region of strong horizontal magnetic field of the PIL, and, thus, are likely associated with a compact sheared magnetic structure elongated along the PIL.

\item The acoustic egression power maps revealed two primary sources of generation of sunquakes, which were associated with places of the strongest photospheric impacts observed in the pre-impulsive phase and early impulsive phase. Thus, we have found an evidence of initiation of helioseismic waves during the pre-impulsive phase prior the HXR impulse, and thus before precipitation of high-energy electrons in the low atmosphere. These two sunquake sources can explain the two-types of helioseismic waves described by \citet{Kosovichev2017}.

\item Analysis of the high-cadence HMI filtergrams suggests that the flare energy release developed in the form of sequential involvement of compact low-lying magnetic loops that are sheared along the PIL. The photospheric emission from different flare points was associated with different HXR bursts. Thus, the particle acceleration process and corresponding photospheric impacts are associated with spatial fragmentation of the flare energy release.

\end{enumerate}

\acknowledgments
The research was supported by the NASA Grants NNX14AB68G, NNX16AP05H and grant of the President of the Russian Federation for the State Support of Young Russian   Science PhDs (MK-5921.2018.2). The observational data are courtesy of NASA/SDO, HMI, GOES, KONUS/WIND and ACS/INTEGRAL science teams.


\bibliographystyle{aasjournal}

\clearpage

\clearpage
\end{document}